\documentclass[reprint, 10pt,a4paper, aps, prl, hidelinks, floatfix]{revtex4-2}
\usepackage{amsmath}
\usepackage{amsfonts}
\usepackage{amssymb}
\usepackage{mathrsfs}
\usepackage[hidelinks]{hyperref}
\usepackage{empheq}
\usepackage{graphicx}
\usepackage{xcolor}
\usepackage{tikz}
\usetikzlibrary{shapes}
\usetikzlibrary{plotmarks}

\newcommand{\br}{\mathbf{r}}
\newcommand{\bv}{\mathbf{v}}

\newcommand{\bE}{\mathbf{E}}
\newcommand{\bB}{\mathbf{B}}
\newcommand{\bF}{\mathbf{F}}

\newcommand{\bR}{\mathbf{R}}
\newcommand{\bg}{\mathbf{g}}
\newcommand{\bp}{\mathbf{p}}

\newcommand{\bomega}{\mathbf{\omega}}
\newcommand{\bw}{\mathbf{w}}

\newcommand{\brho}{\ensuremath{\boldsymbol\rho}}

\begin{document}

\begin{abstract}
Fast rotation can improve the stability and confinement of fusion plasmas. 
However, to maintain a rapidly rotating fusion plasma in steady state, significant energy must be invested in spinning up each incoming fuel ion. 
We show here that, under the right circumstances, collisional cross-field radial fueling can directly transfer drift energy between outgoing and incoming ions without the need for external power recirculation, thereby reducing the energy costs of maintaining the rotation. 
\end{abstract}

\title{Drift-energy replacement effect in multi-ion magnetized plasma}
\date{\today}

%\author{J. M. Rax}
%\affiliation{Andlinger Center for Energy + the Environment, Princeton University, Princeton, New Jersey 08540, USA}
%\affiliation{IJCLab, Université de Paris-Saclay, 91405 Orsay, France}

\author{M. E. Mlodik}
\author{E. J. Kolmes}
\author{N. J. Fisch}
\affiliation{Department of Astrophysical Sciences, Princeton University, Princeton, New Jersey 08544, USA}

\maketitle 

\textit{Introduction:} 
Fast rotation can be desirable in plasma traps for nuclear fusion. 
Rotation often suppresses instabilities \cite{Velikovich1995, Huang2001,Cho2005,Maggs2007, Huneault2019, Haverkort2011,Gueroult2024,Yang2024}.  In centrifugal mirror fusion, rotation  provides axial confinement \cite{Lehnert1971, Bekhtenev1980, Abdrashitov1991, Ellis2005, Teodorescu2010, Romero2012,Endrizzi2023,Ivanov2013}. 
For significant axial confinement, the rotation typically must be supersonic, often with a Mach number much greater than one. 

However, high Mach numbers present a very serious problem for centrifugal fusion devices. 
In steady state operation, new fuel ions  replenish the ions that are burned through fusion. 
For plasma  spinning at a high Mach number,  the energetic ``spin-up cost'' for each new ion can be large, possibly a large fraction of the energy derived from fusion. 
For aneutronic fusion, which occurs at high temperatures, there is an even greater spinning energy cost at a high Mach number, exacerbating an already thin power balance margin \cite{Putvinski2019, Wurzel2022, Kolmes2022, Ochs2022, Magee2023, Ochs2024}. 
However, the same considerations apply to any steady-state, supersonically  rotating fusion plasma. 

Of course, fast plasma  rotation also means that the fusion byproducts are then born with substantial rotation energy. 
When the fusion byproducts are  extracted, their rotation energy might then be captured and converted into electricity, which could offset the energy cost of spinning up the fuel. 
However, converting and re-injecting energy in this way involves substantial inefficiencies. 

However, remarkably, we find that, under certain circumstances, fuel ions may be collisionally exchanged  for the fusion ash (fusion byproducts), while the fuel ion rotation energy is retained in the plasma. 
This surprising result  flows from the constraint that collisions in magnetized plasma do not move charge across field lines.  
In fusion reactions, were the ash to include  neutrons, rotation energy must be lost.  
However, for aneutronic fusion reactions, where both reactants and  byproducts are charged and magnetized, the reactants  can pick up the byproduct rotation energy.
This very fortuitous  {\it drift energy replacement effect} is illustrated in Fig.~1, where ions in the lower slab  have  drift velocities larger than those in the upper slab.  If a helium ion from the upper slab is exchanged with two deuterium ions from the lower slab, which conserves charge, then the helium ion must gain in drift energy, but that energy can be provided exactly by the drift energy of the exchanged deuterium ions.  

\begin{figure}[h]
	\includegraphics[width=0.49\textwidth]{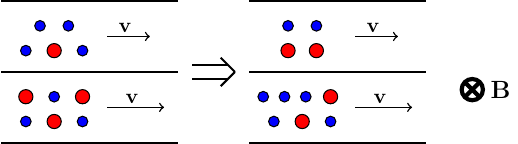}
	\caption{The drift-replacement effect.  For the $E \times B$ drift,  all particles have the same drift velocity perpendicular to the magnetic field, with the  velocities different in the upper and lower  plasma slabs.   Collisions  exchange ions across the slabs, leaving the net charge in each slab  unchanged.  Here, for a D-${}^4$He plasma, the drift energies are completely exchanged.  Red dots represent He (charge 2); the blue dots deuterium.   %  {    \color{red} Shouldnt the velocities in the upper slab both be the same and larger than those in the lower slab?   }.   
	}
	
\end{figure}

~\newline
\textit{Stating the Cost Issue:} 
Before uncovering the conditions under which and the extent to which the drift energy replacement effect occurs, let us state precisely the cost issue.  
%  Consider a box of plasma at temperature $T$. %  with perfect energy confinement. 
The energy gained per fusion event is $\Delta \epsilon =  \alpha \left( \epsilon_{fus} + \epsilon_{kin} \right) - \beta \epsilon_{kin}$, where $\epsilon_{fus}$ is the fusion energy yield, $\epsilon_{kin}$ is total thermal energy residing in particles participating in a fusion event, $\alpha \leq 1$ is the fusion energy conversion efficiency, and $\beta \geq 1$ is the energy cost of fueling a particle and heating it up to the temperature of the fusing plasma. For viable fusion, $\Delta \epsilon \geq 0$, or  
\begin{gather}
\frac{\beta}{\alpha} - 1 \leq \frac{\epsilon_{fus}}{\epsilon_{kin}}.
\label{eqn:efficiencyBasic}
\end{gather}
Now suppose that plasma in the box is drifting. Then the drift energy must be taken into account.  
As it turns out, how to do that  depends on  how the plasma is fueled. 

Suppose the plasma is fueled via the injection of (stationary) neutral pellets. 
Upon ionization, a neutral picks up $m \bv_d^2/2$ drift energy and the same energy in Larmor gyration (perpendicular heating), based on instantaneous momentum conservation. % Specifically, a newly created ion acquires drift velocity $\bv_d$ and thermal velocity $-\bv_d$. 
Both drift energy and heating are at the expense of the electric field energy, which is lowered through the  induced charge separation. 

% If $\text{Ma}$ is the appropriately averaged 

For Mach number $\text{Ma}$ of the plasma flow,  the drift kinetic energy is $\epsilon_{dr} = \text{Ma}^2 \epsilon_{kin}$, so that  % and the energy balance is 
\begin{gather}
	\Delta \epsilon =  \alpha \left( \epsilon_{fus} + \epsilon_{kin} + \epsilon_{dr} \right) - \beta \left( \epsilon_{kin} + \epsilon_{dr} \right),
	\label{eqn:energyBalancePellet}
\end{gather}
producing the more stringent requirement on $\alpha$ and $\beta$ 
\begin{gather}
	\frac{\beta}{\alpha} - 1 \leq \frac{\epsilon_{fus}}{\epsilon_{kin}} \frac{1}{1 + \text{Ma}^2}.
	\label{eqn:efficiencyPelletFueling}
\end{gather}
The drift energy must then be externally recirculated (removed from fusion products, converted to electricity, and then used to sustain the electric field).   
% so they are subject to the same kinds of power-conversion inefficiencies associated with heating the fuel ions. 
Similarly, if the pellet is instead injected at high velocity,  the conclusions are largely the same, due to the launch energy cost. 

% add pellet cartoon? 
%This is a less serious issue for DT fuel, where $\epsilon_{fus} = 17.6$ MeV and $\epsilon_{kin} \sim 10-20$ keV. However, the limit on $\text{Ma}$ set by Eq.~(\ref{eqn:efficiencyPelletFueling}) is stricter for advanced fuels ($DD$, $D{}^{3}He$, $pB$, etc.) because they require higher temperatures and have somewhat lower fusion power output. 
%This requirement can become substantially more serious when one considers the additional inefficiencies from imperfect energy confinement. 

It is thus both surprising and fortuitous that collisional cross-field fueling can facilitate a direct transfer of the rotational kinetic energy from outgoing to incoming ions. 
If the drift energy exchange is one-to-one, the energy balance that leads to Eq.~(\ref{eqn:efficiencyBasic}) is unaffected.

%per fusion event as in the basic case:
%\begin{gather}
%\Delta \epsilon =  \alpha \left( \epsilon_{fus} + \epsilon_{kin} \right) - \beta \epsilon_{kin}.
%\label{eqn:energyBalanceCollisionalMatch}
%\end{gather}
%In other words, the ``engineering efficiencies'' captured by $\alpha$ and $\beta$ do not apply to the drift energy because there is a natural, high-efficiency transfer during the refueling process. 
%This removes the restriction on the Mach number of the plasma flow. 

% The rest of this Letter is concerned with the details of how this kind of high-efficiency transfer can take place, and when it is possible. 

%~\newline
%\textit{Collisional fueling. Single-particle model of drift-replacement effect in slab geometry.} 

~\newline
\textit{The Drift Energy Replacement Effect:}
Consider a plasma slab immersed in a uniform magnetic field $\bB = B \hat z$, in Cartesian $(x,y,z)$ coordinates with unit vectors $\hat x$, $\hat y$, and $\hat z$. 
Suppose the plasma is homogeneous in the $\hat y$ direction, with plasma parameters such as the density $n$ varying only in the $\hat x$ direction. 
%It is productive to consider the effects of an individual collisional interaction, and to consider the correlation between its effects on each particle as follows. 
%Because the correlation is the same (up to leading order) for all collisions between two particles with given identities, the same correlation between the change in drift energy of different species will hold. 
%
Consider two particles interacting at point $\br$, where
\begin{gather}
\br = \bR_j + \brho_j.
\end{gather}
Here $\bR_j$ is position of gyrocenter of particle $j$, $\brho_j$ is the vector between gyrocenter and actual particle position, and we have taken the Larmor radii to be large compared with any distance between the interacting particles. 
The velocity of either particle can be expressed as 
\begin{gather}
\bv_j = \bv_{dj} (\bR_j) + \bw_j (\brho_j) + v_{||j} \hat{\mathbf{b}} . 
\end{gather}
Here $\bv_{dj}$ is the drift velocity, $\bw_j$ is the velocity associated with gyromotion, $v_{||j} \hat z$ is the motion parallel to the magnetic field, and $\hat{\mathbf{b}} \doteq \bB / B = \hat z$. 
The gyromotion velocity can be expressed as
\begin{gather}
\bw_j = \Omega_j \brho_j \times \hat{\mathbf{b}}.
\label{eqn:gyromotion}
\end{gather}
Here $\Omega_j = q_j B/m_j$ is the gyrofrequency of particle $j$, where $q_j$ and $m_j$ are the charge and the mass of the particle, respectively.
Conversely,
\begin{gather}
\brho_j = - \frac{\bw_j \times \hat{\mathbf{b}}}{\Omega_j}.
\end{gather}
Note that collisions between particles are instantaneous compared with the Larmor gyration time of each particle. Therefore, the electromagnetic part of the momentum is not changed during the collision itself, and the total kinetic momentum of two particles is conserved. 

Let $\delta \bp$ be the kinetic momentum transferred from the second to the first particle during the collision and $\delta \bp_j$ be the change of kinetic momentum of particle $j$ (i.e. $\delta \bp_1 = \delta \bp$, $\delta \bp_2 = - \delta \bp$). Then 
\begin{gather}
\delta \bv_{dj} + \delta \bw_j = \frac{\delta \bp_j}{m_j};
\end{gather}
\begin{gather}
\delta \bR_j + \delta \brho_j = 0.
\end{gather}
% The equations above assume that the drift velocity profiles $\bv_{dj} (\bR_j)$ are the same before and after collision. 
%
Using Eq.~(\ref{eqn:gyromotion}), we have  $\delta \bw_j = \Omega_j  \delta \brho_j \times \hat{\mathbf{b}}$, 
so that:
\begin{gather}
\delta \bv_{dj} - \Omega_j \delta \bR_j \times \hat{\mathbf{b}} = \frac{\delta \bp_j}{m_j}.
\label{eqn:flowVelocityChangeFull}
\end{gather}
%Then the change in total flow energy is 
%\begin{align}
%\delta W_{flow} = \sum_{j=1,2} m_j \left[ \bv_{dj} \delta \bv_{dj} + \frac{1}{2} (\delta \bv_{dj})^2 \right].
%\end{align}
%

Let us write $\delta \bv_{dj} = \bomega_j \cdot \delta \bR_j$.   We will assume constant $\bomega_j$,  which is exact if the shear is constant and it holds to leading order  in the ion gyroradius  compared to the shear length scale.  (If the flow is incompressible, $\bomega_j$ is related to the vorticity.)
In the geometry  here, $\bomega_j = |\bomega_j| \hat{y} \hat{x}$. 
The equations of motion in the perpendicular ($x$-$y$) plane can then be solved separately from those in $z$-direction,
% Therefore, the latter equations can be ignored in this analysis. 
with the $\hat x$ and $\hat y$ components of $\delta \bR_j$  satisfying
\begin{gather}
\begin{pmatrix}
0 & - \Omega_j \\
\Omega_j + |\bomega_j| & 0
\end{pmatrix}
\delta \bR_j = \frac{\delta \bp_j}{m_j}.
\end{gather}
The solution for $\delta \bR_j$ is
\begin{gather}
\delta \bR_j = 
\begin{pmatrix}
0 & \frac{1}{\Omega_j + |\bomega_j|}  \\ - \frac{1}{\Omega_j} 
& 0
\end{pmatrix}
\frac{\delta \bp_j}{m_j}.
\label{eqn:gyrocenterJump}
\end{gather}
The structure of this correlation draws out much of the essential physics described in this paper (playing a similar role to the correlations used in the alpha-channeling problem \cite{Fisch1992}). 
Using this, we can show that the change in drift velocities is 
\begin{gather}
\delta \bv_{dj} = \hat{y} \frac{|\bomega_j|}{\Omega_j + |\bomega_j|} \frac{\delta p_{jy}}{m_j}.
\end{gather}
The change of drift velocity energy due to a collision can be written as
\begin{align}
\delta W_{drift}  = \sum_{j=1,2} \bv_{dj} \cdot m_j \delta \bv_{dj} + \frac{1}{2} \sum_{j=1,2}  m_j \left(\delta \bv_{dj}\right)^2,
\end{align}
or, alternatively,
\begin{align}
\delta W_{drift} & = \delta p_y \left[  v_{d1y} \frac{|\bomega_1|}{\Omega_1 + |\bomega_1|} -  v_{d2y} \frac{|\bomega_2|}{\Omega_2 + |\bomega_2|} \right] \nonumber\\  & + \frac{1}{2} \left[ m_1 \left(\delta \bv_{d1}\right)^2 + m_2 \left( \delta \bv_{d2} \right)^2 \right].
\end{align}

In the case of small shear (i.e $|\bomega_1| \ll \Omega_1, |\bomega_2| \ll \Omega_2$), the change in drift energy is
\begin{gather}
\delta W_{drift} = \delta p_y \left( \frac{v_{d1y} |\bomega_1|}{\Omega_1} - \frac{v_{d2y} |\bomega_2|}{\Omega_2} \right).
\label{eqn:flowEnergyChange}
\end{gather} 
The change of drift energy associated with particle transport of particles of species $1$ represents the correlation between $\delta R_{1x}$ (jump in gyrocenter position of particle $1$ due to collision) and $\delta W_{drift}$.  
Using the expression for $\delta R_{1x}$ from Eq.~(\ref{eqn:gyrocenterJump}), and taking the drift velocities $\bv_{dj}(\br)$ to be the same for both species (as is the case for the $\bE\times\bB$ drift), this correlation can be put as
\begin{gather}
\frac{\delta W_{drift}}{\delta R_{1x}} = m_1 v_{dy} | \bomega | \left(1 - \frac{\Omega_1}{\Omega_2}\right) \left[ 1 + \mathcal{O} \left(\frac{|\bomega|}{\Omega}\right) \right].
\label{eqn:flowEnergyCorrelation}
\end{gather}

Eqs.~(\ref{eqn:flowEnergyChange}) and (\ref{eqn:flowEnergyCorrelation}) represent the {\it drift-energy replacement effect}, the major result of this Letter.  These equations demonstrate that collisions between species $1$ and $2$  drive in correlation both the cross-field transport and the transfer of drift energy between the two species. 
Eq.~(\ref{eqn:flowEnergyCorrelation}) shows that, for small enough shear in drift velocity,  regardless of the details of collision operator, the drift velocity profiles, 
% (which would have entered as statistics of $\delta \bp$), 
and the extent of collisional transport,   % (as long as shear across gyroradius is small). 
the change in total rotation energy is $\left(1-\Omega_1/\Omega_2\right)$ times the rotation energy acquired by the particle of species $1$.

To understand Eq.~(\ref{eqn:flowEnergyCorrelation}), consider some limiting cases. If $|\omega| = 0$, which is equivalent to $\bv_{dj} = const$ (uniform drift), then $\delta W_{drift} = 0$. 
In other words, if plasma moves as a whole without shear, there is no change in drift energy due to collisional transport. 
If $\Omega_1 = \Omega_2$ (both types of particles have the same gyrofrequency), then $\delta W_{drift} = 0$. 
In other words, the drift energy gained by incoming ions is matched by the drift energy given up by the outgoing ions, and no interaction with other energy reservoirs, such as potential energy, is required. 

%Put later
%The ordering parameter which determines the accuracy of the energy cost  is $\mathcal{O} (|\omega|/\Omega) = \mathcal{O} (\text{Ma} \, \rho/L_{\perp})$, where $\text{Ma}$ is the ion Mach number, $\rho$ is ion gyroradius, and $L_{\perp}$ is the characteristic length scale of the plasma in the direction perpendicular to $\bB$. 
%The derivation of the drift-energy replacement effect is shown up to first order in the parameter $\text{Ma} \, \rho/L_{\perp}$.
%Note that the heating associated with this process is higher order in $\delta v_d / v_d$. 
%Therefore, the heating is proportional to the fueling rate. 
%Thus, the  heating is limited for ion density profiles  close to  equilibrium, as would arise naturally in steady state fueling.

\begin{figure}[h]
	\begin{tikzpicture} [scale = 0.6]

	%	\draw [line width = 3, fill = white] (-8,-5)
	%	rectangle (6,4);
	
	\draw [line width = 1, -, color = red, dashed] (-4.8, 3.0) to (4.5, 3.0);
	%	\node [align = left] at (0, 2+0.5) {\large $\Gamma_b$}; 		
	%	
	
	\draw [line width = 2, -, color = blue] (4.5, -3.0) to (-4.8, -3.0);
	%	\node [align = left] at (0, -2+0.5) {\large $\Gamma_a$}; 			
	
	%	\node [align = left] at (-3.7, 2) 
	%	{\large $n_b(x_1)$};
	%	
	%	\node [align = left] at (-3.7, -2) 
	%	{\large $n_a(x_1)$};
	%	
	%	\node [align = left] at (3.7, 2)
	%	{\large $n_b(x_2)$};
	%	
	%	\node [align = left] at (3.7, -2)
	%	{\large $n_a(x_2)$};
	
	\draw [line width = 1, ->, color = black, dashed] (-1.9, -2.5) to (-1.9, 2.5);
	\node [align = left] at (-1.9+0.6, 0) {\large 	$\Gamma_B$}; 			
	
	\draw [line width = 3, ->, color = black, dashed] (4, 2.5) to (4, -2.5);
	\node [align = left] at (4-1.6, 0) {\large 	$\Gamma_p = 5 \Gamma_B$}; 			
	
	\node [align = left] at (-4.5+0.6, 0.7+0.5) {$v_y(x)$}; 				
	\draw [line width = 1, ->, color = black] (-4.5, 0.7) to (-4.5+1.2, 0.7);
	\draw [line width = 1, ->, color = black] (-4.5, 0.0) to (-4.5+1.0, 0.0);
	\draw [line width = 1, ->, color = black] (-4.5, -0.7) to (-4.5+0.8, -0.7);

	\node [align = left] at (5.6, 3.0)
	{\large core};
	
	\node [align = left] at (5.6, -3.0)
	{\large edge};
	
	\draw [line width = 2, fill = white] (5.3, 0) circle (0.3); 
	%					\draw [line width = 3, fill = black] (4.5, 2.5) circle (0.1 cm); 
	\node [align = left] at (5.3+0.8, 0) {\large $\textstyle{\mathbf{B}}$};
	
	\draw [line width = 2] ({5.3+0.3*cos(135)}, {0+0.3*sin(135)}) to ({5.3+0.3*cos(315)}, {0+0.3*sin(315)});
	\draw [line width = 2] ({5.3+0.3*cos(45)}, {0+0.3*sin(45)}) to ({5.3+0.3*cos(225)}, {0+0.3*sin(225)});
	
	% individual velocities	
	
	\draw [line width = 1, ->, color = black] (-1.3, 1.9) to (-1.3+1.2, 1.9);
	\node [align = left] at (-1.3+0.7, 1.9+0.4) {$v_{By,fin}$}; 			
	
	\draw [line width = 1, ->, color = black, dotted] (-1.3, -2.1) to (-1.3+0.8, -2.1);
	\node [align = left] at (-1.3+0.7, -2.1+0.4) {$v_{By,in}$}; 			
	
	\draw [line width = 1, ->, color = black,dotted] (2.2, 1.9) to (2.2+1.2, 1.9);
	\node [align = left] at (2.2+0.7, 1.9+0.4) {$v_{py,in}$}; 			
	
	\draw [line width = 1, ->, color = black] (2.2, -2.1) to (2.2+0.8, -2.1);
	\node [align = left] at (2.2+0.7, -2.1+0.4) {$v_{py,fin}$}; 			
	
	%	\node [align = left] at (-3.7, -4)
	%	{\large $x = x_1$};
	%	
	%	\node [align = left] at (3.7, -4)
	%	{\large $x = x_2$};

	\end{tikzpicture}
	\caption{Cross-field boron fueling   while replacing protons.}
	\label{multiple_ions}	
\end{figure}
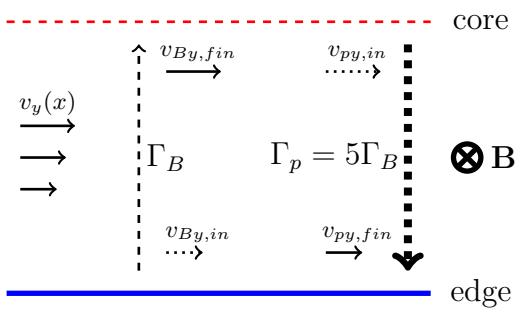

This can substantially reconfigure the flow of energy as particles move in and out of the system. 
For example, take boron as species $1$ and protons as species $2$, as  shown in Fig.~\ref{multiple_ions}.  
Then $5/11$ of the boron rotational energy is provided by the protons, which are deconfined, leaving only  $6/11$ to be taken from the potential energy. 
Notably, it is also possible to extract drift energy.  
If $\Omega_1 > \Omega_2$, then $\delta W_{drift}$ is negative, so outgoing ions have smaller gyrofrequency than incoming ions.  
Then outgoing ions not only give their rotational energy  to the incoming ions, they also recharge the potential energy of the field.

\textit{Multiple Ion Drift Energies:}
The discussion thus far considered  only two species of interacting ions. 
For more than two species, high-efficiency energy transfers can be accomplished even without precise matching of the gyrofrequencies. 
For an arbitrary number of interacting ion species, the change in drift energy can be written as: 
\begin{align}
\delta W_{drift} &= \sum_j m_j \bv_{dj} \cdot \delta \bv_{dj} \bigg[ 1 + \mathcal{O} \bigg( \frac{|\delta \bv_{dj}|}{|\bv_{dj}|} \bigg)  \bigg] \\
%&= \sum_j v_{djy} |\omega_j| \frac{\delta p_{jy}}{\Omega_j} 
%\nonumber \\ &\hspace{40 pt} \times 
%\bigg[ 1 + \mathcal{O} \bigg( \frac{|\omega_j|}{\Omega_j} \bigg) + \mathcal{O}\bigg( \frac{|\delta \bv_{dj}|}{|\bv_{dj}|} \bigg) \bigg] \\
&= \sum_j v_{dj} |\omega_j| m_j \delta R_{jx} 
%\nonumber \\ &\hspace{40 pt} \times 
\bigg[ 1 + \mathcal{O} \bigg( \frac{|\omega_j|}{\Omega_j} \bigg) + \mathcal{O}\bigg( \frac{|\delta \bv_{dj}|}{|\bv_{dj}|} \bigg) \bigg] . 
\end{align}
In the limit where all species have the same drift velocity profile, this means that (to leading order) $\delta W_{drift}$ vanishes any time the inward and outward mass flux is equal. 
This is equivalent to matching gyrofrequencies in the two-species case because, in an interaction between two species, $Z_1 \delta R_{j1} + Z_2 \delta R_{j2} = 0$.
Thus, for example, in pB11 centrifugal fusion devices, utilizing  $E \times B $ drifts,  there would be complete drift energy replacement for both protons and boron  replacing alpha particle ash.

~\newline
\textit{Drift Energy Release in a Fusion Event:}
Interestingly, and closely related to the drift energy replacement effect, depending on the physical mechanism of the drift, and the species involved, drift  energy can be released in a fusion event in magnetized plasma. 
Consider a force $\bF_s = - \nabla \Phi_s$. The gyroradius vector of a particle $s$ is
\begin{gather}
\brho_s = - \frac{m_s \bw_s \times \hat{\mathbf{b}}}{q_s B}.
\end{gather}
The potential energy of particle $s$ relative to the location of the fusion event is 
\begin{gather}
W_s = \bF_s \cdot \brho_s.
\end{gather}
Given that $\bw_s = \bv_s - \bv_{ds}$,
\begin{gather}
W = \sum_{s} W_s = - \sum_s \frac{m_s \left(\bv_s - \bv_{ds}\right) \times \hat{\mathbf{b}} \cdot \bF_s }{q_s B}.
\end{gather}
Alternatively, if $\bp_s = m_s \bv_s$, then
\begin{gather}
W = \sum_{s}\frac{m_s \bv_{ds} \times \hat{\mathbf{b}} \cdot \bF_s}{q_s B} - \sum_{s} \frac{\bp_s \times \hat{\mathbf{b}} \cdot \bF_s}{q_s B}.
\end{gather}
%After cycling triple product, we can get
%\begin{gather}
%W = \sum_s \frac{\bF_s \times m_s \bv_{ds}}{q_s B} \cdot \hat{\mathbf{b}} - \sum_s \frac{\bF_s \times \bp_s}{q_s B} \cdot \hat{\mathbf{b}}.
%\end{gather}
%
%
If $\bv_{ds}$ is caused by total force $\bF_{tot,s}$, then
\begin{gather}
W = - \sum_s \frac{\bF_s \cdot \bF_{tot,s}}{m_s \Omega_s^2} - \sum_s \frac{\bF_s \times \bp_s}{q_s B} \cdot \hat{\mathbf{b}}.
\end{gather}

If the $\bE \times \bB$ drift is the only drift, then the second term does not change in a fusion event due to momentum conservation. The first term becomes $\sum_s m_s E^2/B^2$ and is constantt (to the extent that the fusion event conserves total mass). Therefore, there is no change in electric potential in a fusion event in an $\bE \times \bB$ drifting  plasma.

The situation is very different if the force is proportional to mass, for example, if  $\bF_s = m_s \bg$, then
\begin{gather}
W = - \sum_s\frac{m_s^3}{q_s^2} \frac{g^2}{B^2} + \sum_s \frac{\bp_s}{\Omega_s} \cdot \left( \bg \times \hat{\mathbf{b}}\right).
\end{gather}
Now, in general, the potential energy  $W$ changes in a fusion event. 
For example, in the pB11 reaction, the change of the first term is $\Delta W_1 = +6.24 \left(m_p^3/q_p^2\right) \left(g^2/B^2\right)$, while the change of the second term is 
\begin{gather}
\Delta W_2 = \left[\bp_p \left( \frac{1}{\Omega_\alpha} - \frac{1}{\Omega_p}\right) + \bp_B  \left( \frac{1}{\Omega_\alpha} - \frac{1}{\Omega_B}\right) \right] \cdot \left(\bg \times \hat{\mathbf{b}}\right).
\end{gather}
Alternatively,
\begin{gather}
\Delta W_2 = - \left( \frac{1}{2} \bp_p  - \frac{1}{22} \bp_B \right) \cdot \frac{ \left(\bg \times \hat{\mathbf{b}}\right) }{\Omega_p} .
\end{gather}
Thus, there is a release of potential energy as a result of the fusion event, unless $\Delta W_1 + \Delta W_2 = 0$.

Of practical interest in fusion devices is the centrifugal force, rather than the gravitational force, which is also proportional to mass. 
In a rapidly rotating system where mass-dependent inertial drifts are important, the redistribution of the gyrocenters after a fusion event  changes the total potential energy of the participating particles. 
The mechanism underlying this is much the same as the one behind the drift-energy replacement effect: interactions between particles at a given point can shift their gyrocenters up or down a potential gradient. 

In the case of the pB11 reaction undergoing fusion in a centrifugal mirror machine, the mass-weighted centrifugal force on the fusion reactants is greater than the mass-weighted force on the fusion byproducts, so the rotational energy of the fusion reactants is greater than that of the fusion byproducts.   
Thus, drift energy is released in the pB11 fusion reaction.  
In fact, this energy release is greatest when the centrifugal fusion reactor is charged positive (the usual case), for then the centrifugal forces are in the same direction as the electric forces, thus enhancing the drift energy difference between the reactants and the byproducts.  The energy release settles in part the account of drift energy to be replaced. It is supplementary in fact to the collisional drift energy replacement effect.    To the extent that the fusion byproducts do not have quite the drift energy of the fusion reactants, then the  collisional drift energy replacement effect cannot replace the full reactant drift energy.  However, the sum of the two effects does replace the full reactant drift energy.  

~\newline
\textit{Summary and discussion.}
 The power cost of refueling ions can render uneconomical fusion devices requiring high plasma drift velocities.
Although fusion products might be recovered with that drift energy,  the recovery and re-injection of the energy (as new fuel ions need to be ``spun up") can be a source of significant inefficiency. 

This Letter identifies the {\it drift-energy replacement effect}, a new and fundamental transport effect that can obviate the need to recover the drift energy of fusion ash.
% which can -- under the right circumstances -- make it possible to avoid this problem, and thereby avoid limitations on the viable Mach number of a fusion device. 
Surprisingly, when ions of different species move via collisions across magnetic field lines in opposite directions, incoming ions can automatically take all  or part the drift energy directly out of the outgoing ions. 

Near equilibrium, this process represents a particular and fortuitous case of multiple-ion-species redistributions \cite{Bonnevier1966, Krishnan1983, Geva1984, HelanderSigmar, Zhdanov, Kagan2012, Kagan2014,Helander2017, Kolmes2018,Mlodik2020, Angioni2021, Mitra2021, Kolmes2021,Rubin2021}. 
For plasma near collisional equilibrium, irreversible heating associated with particle transport can take place, but is small if the transport is not fast.
Since the ion rearrangement itself does not produce heating, it is reversible;  with reversed boundary conditions or forces, the net result of ion exchanges can be reversed.
The interaction between incoming and outgoing ions is not specific to collisions, so the calculations in this Letter are applicable also to wave-mediated transport, so long as the wave does not impart net momentum to  ions and the momentum transfer can be regarded as instantaneous.
%

%In the two-species case, if the gyrofrequencies of the incoming and outgoing ions are the same, the outgoing ions transfer exactly enough energy to the incoming ions to get them to the drift velocity. 
%More generally, high-efficiency energy transfer is possible when the total cross-field collisional mass flux of incoming and outgoing particles is equal. 
The energy transfer is possible at near-perfect efficiencies, and is reversible, but only if the conditions outlined here are met.
In cases where the transfer does not  meet the drift energy needed, the difference between the drift energies of incoming and outgoing ions would be provided by the potential energy driving the drift (the electrostatic potential, in the case of the $\bE\times\bB$ drift). 

This difference can also be negative if more drift energy is removed than replaced; in that case, power can be drawn from the circuit providing for the rotation.
Also, note that the drift energy replacement cannot be perfect in neutronic reactions, as the balance of incoming and outgoing particles depends only on those species which participate in the collisional transport; neutrons leave promptly, so any drift energy they carry would not be recoverable within the plasma.

%In particular, in pulsed systems, where the influx and outflux of particles is not , the release of drift energy over the course of a pulse cycle. 
%There are also distinctions to be drawn between systems operated in steady state and those that are pulsed. 
%In a system not operating in full steady state, t
%Irreversible heating associated with this transport can take place, but is small so long as the transport is not too fast. 

The drift-energy replacement effect shows that, in fusion plasmas with large drift flows, the fueling mechanism can significantly impact the power flow, especially if advanced fuels are considered. 
Although we focus on nuclear fusion, the replacement effect uncovered here applies also to other technologies involving rapid rotation, such as rotating plasma mass filters \cite{Lehnert1971, Krishnan1983, Geva1984, Dolgolenko2017, Zweben2018, Gueroult2018ii}. % Kaganovich2020}. 

Apart from its inherent academic interest, the drift-energy replacement effect identified here has very practical interest; it represents the very rare case in which collisional effects in particular render economic nuclear fusion to be easier rather than harder. 
The cost of spinning up new fuel ions could very well have proved prohibitive for any steady-state fusion reactor based on the centrifugal confinement of fuels for aneutronic fusion, since, for the needed sonic or supersonic rotation, these very hot plasmas would feature very large invested rotation energy.  
At the same time, reactor designs based on aneutronic reactions tend to have very small margins of net energy produced compared to invested energy. 
Even for more conventional lower-temperature fuels, the spin-up cost could be significant. 
But if this spinning-up can be accomplished ``for free," with a direct transfer of rotation energy from the outgoing fusion products, then this cost will not be as formidable as might have been expected. 

% EJK: I don't think I understand what is meant by the following passage. Probably good to discuss at some point, but I'm removing it for now. 
%If further improvements in delivering flow energy to fuel ions are required, then fueling mechanism requires to deliver net momentum to ions as a whole, at rate which is at least comparable to the rate of momentum exchange between ion species. Note, however, that this is not a trivial requirement, because multi-ion plasma transport implies timescales separation \cite{Mlodik2020,Kolmes2021}, and dominant collision-related term, ion-ion friction, doesn't change total kinetic momentum.

%\textcolor{blue}{EJK: A question about how to think about this. In some cases, the imbalance in $\delta W_{flow}$ results in discharging the perpendicular electric field. It seems clear that this leads to inefficiency when we need to recharge the field. But what about the cases where it \textit{charges} the field? Should we expect to be able to recover this energy at high efficiencies? Right now the paper presents $\delta W_{flow} = 0$ as the ``good" case, but if we can efficiently recover any extra energy that ends up in $\bE$, should we instead be saying we want $\delta W_{flow} \leq 0$? } 

\textit{Acknowledgments.} The authors thank I. E. Ochs, T. Rubin,  J.-M. Rax, and V. R. Munirov for useful conversations. This work was supported by ARPA-E Grant DE-AR0001554. 
EJK also acknowledges the support of the DOE Fusion Energy Sciences Postdoctoral Research Program, adminstered by the Oak Ridge Institute for Science and Education (ORISE) and managed by Oak Ridge Associated Universities (ORAU) under DOE Contract No. DE-SC0014664. 
\bibliographystyle{apsrev4-2} 
\providecommand{\noopsort}[1]{}\providecommand{\singleletter}[1]{#1}%

\end{document}